\documentclass[twocolumn,showpacs,preprintnumbers,amsmath,amssymb,aps]{revtex4}
\usepackage{graphicx}% Include figure files
\usepackage{dcolumn}% Align table columns on decimal point

\usepackage{color}

\newcommand{\be}{\begin{equation}}
\newcommand{\ee}{\end{equation}}
\newcommand{\bea}{\begin{eqnarray}}
\newcommand{\eea}{\end{eqnarray}}
\newcommand{\ds}{\displaystyle}

\newcommand{\kk}{{\bf k}}

\newcommand{\zzeta}{{\boldsymbol\zeta}}
\newcommand{\Angstrom}{{\buildrel _{\circ} \over {\mathrm{A}}}}

\begin{document}
%\title{Implementation of a Stimulated Raman Amplifier/Compressor in Plasma}
\title{Beam debunching due to ISR-induced energy diffusion}
\author{Nikolai A. Yampolsky and Bruce E. Carlsten}
\affiliation{Los Alamos National Laboratory, Los Alamos, New Mexico, 87545, USA}
%\date{\today}

\begin{abstract}
One of the options for increasing longitudinal coherency of X-ray free electron lasers (XFELs) is their seeding with microbunched electron beam.
Several schemes leading to significant amplitude of the
beam bunching at X-ray wavelengths were recently proposed. All these schemes rely on beam optics having
several magnetic dipoles. While the beam passes through a dipole, its energy spread increases due to
quantum effects of synchrotron radiation. As a result, the bunching factor at small wavelengths reduces
since electrons having different energies follow different trajectories in the bend. 
We rigorously calculate reduction in the bunching factor due to
incoherent synchrotron radiation while the beam travels in arbitrary beamline. We apply general results to estimate reduction of harmonic current in 
common schemes proposed for XFEL seeding.
\end{abstract}

\maketitle

\section{Introduction}

Free electron laser (FEL) can be scaled to generate narrowband radiation in a wide frequency range. Currently existing FELs cover
bandwidths from THz to hard X-rays. Conventional method for decreasing the FEL bandwidth implies placing the undulator inside the optical cavity. However, this approach
fails for X-ray FELs (XFELs) due to lack of high quality reflecting mirrors at these frequencies.
As a result, modern XFELs operate in a single pass self-amplified spontaneous emission (SASE) regime in which
broadband shot noise due to discreetness of the electron beam is amplified within the FEL bandwidth \cite{SASE}. The bandwidth of these XFELs can be reduced through their seeding with a narrowband signal
rather than with a white noise.

Growing FEL mode couples radiation with particle motion. Therefore, XFEL can be seeded either with a narrowband radiation \cite{optical_seeding,selfseeding} or with electron beam modulated at the X-ray
wavelength \cite{HGHG}. The second option looks attractive since electrons are charged particles and they can interact with electromagnetic fields unlike radiation which only weakly interacts with
materials at X-ray frequencies. Several schemes predicting significant bunching at high harmonics of available coherent light sources were recently proposed \cite{EEHG,CHG,EEHG-EEX}. All these
schemes utilize magnetic bends for manipulations with modulated electron bunch. Electrons passing through bend emit synchrotron radiation.
At high energies electrons emit radiation at high frequencies and quantum effects should be included to find correct electron energy loss.
Incoherent synchrotron radiation (ISR)
has nearly 100\% frequency spread which results in the electron energy diffusion when quantum effects are accounted since electron energy loss can be described within random
walk model. Electron energy diffusion translates into diffusion along longitudinal position since bends are dispersive elements (electrons with different energies travel along different trajectories).
As a result, prebunched electron beam smears out when it passes through bends and the bunching factor reduces.

The effect of the beam debunching becomes stronger for FELs seeded at smaller wavelengths. First of all, energetic electron bunch is required to generate short wavelength FEL
radiation which increases ISR-induced energy diffusion in the beamline bends. At the same time, the associated longitudinal spread should be smaller to cause 
significant debunching of shorter wavelength modulation. These two effects combined result in stronger smearing of harmonic current of short wavelength XFELs. Therefore,
there is a technological limit on the shortest wavelength bunching which can be created in various schemes for generating high harmonic content.
This effect was estimated for various XFEL seeding schemes \cite{EEHG,CHG_Ratner,Litvinenko} but these estimates remain qualitative or numerical and do not allow for detailed trade-off studies.
It is the purpose of this paper to study this effect rigorously and find accurate quantitative estimate on degradation of harmonic current in various beamlines.

\section{Qualitative estimate for smearing of harmonic current}

First, we estimate the parameters region in which smearing of harmonic current due to ISR-induced energy spread is significant. 
We consider any FEL seeding scheme for creating longitudinal bunching. Consider the last dipole of the beamline optics which recovers imposed modulation as longitudinal bunching.

Quantum effects in ISR at high energies result in energy diffusion as described  in Ref.~\cite{ISR:energy}
\be
\label{dE}
\left\langle {\Delta E^2}\right\rangle=2Ds,\;\;\;\;\;D=\frac{55}{{48\sqrt{3}}}\frac{\hbar e^2 c}{\rho^3}{\gamma^7},
\ee
where $D$ is the energy diffusion coefficient,
$\rho=\beta\gamma mc/(eB)$ is the electron gyroradius in the magnetic field $B$; $\hbar$ is the reduced Planck constant, and $s$ is the path length. Two identical 
electrons entering the dipole exit at different longitudinal positions since their energies become different inside the dipole due to ISR and the dipole energy dispersion transforms this energy
difference into difference in longitudinal positions. As a result, bunched beam smears out.
We approximate this effect assuming energies of two electrons to be constant along the bend but to be different by the overall induced
energy spread due to ISR, $\Delta E=\sqrt{2Ds}$.
The path length of an electron inside the magnet of length $L$ is equal to
\be
s=\rho\arcsin\frac{L}{\rho}.
\ee
Then the path difference of electrons having different energies is 
\be
\Delta s=\frac{\partial s}{\partial \rho}\frac{\partial \rho}{\partial E} \Delta E\approx-\frac{L}{3}\alpha^2\frac{\Delta E}{E}
\ee
This path difference results in smearing of the imposed modulation. Smearing is significant if the path difference is on the order
of half wavelength of modulation. Then one obtains the qualitative estimate for parameters region in which  ISR-induced debunching is not significant.
\be
\label{qualitative}
\frac{55}{54\sqrt{3}}\frac{\hbar e^2}{m^2c^3}\frac{\alpha^7\gamma^5}{\lambda^2}=
90.6\frac{\alpha_0^7[{\rm deg}] }{\lambda^2[{\buildrel _{\circ} \over {\mathrm{A}}}]}\left(\frac{E}{10 GeV}\right)^5\ll1.
\ee

Presented estimate shows that beam bunching for hard X-ray FELs ($\lambda\sim 1 \buildrel _{\circ} \over {\mathrm{A}}$, $E\sim 10 {\rm GeV}$) requires the use of optical elements with weak
bends having sub-degree bend angles. However, this estimate does not predict how fast the bunching degrades along the bend. In particular, it is not clear how fast the bunching drops to zero and how
far one can go beyond this limit without dramatic loss in bunching amplitude.

\section{Smearing of harmonic current in arbitrary beamline}
\label{sec:Vlasov}

\subsection{Vlasov equation}
Any electron beam can be described as an ensemble of electrons occupying some phase space volume. This ensemble can be described with the phase space distribution function $f(\zzeta)$, where $\zzeta$
is the 6D phase space coordinate of each electron
\be
\label{zeta}
{\bf \zzeta}=(x, p_x, y, p_y, \Delta t, -\Delta E),
\ee
where $x$ and $y$ are the transverse electron coordinates in respect to the reference trajectory, $p_x$ and $p_y$ are the corresponding momenta, $\Delta t$ is the deviation of arrival time
to position $s$ along the beamline, $\Delta E$ is the deviation of particle energy from the average bunch energy. 

The bunch interacts with electro-magnetic fields while it travels along the beamline. These forces satisfy Maxwell equations which indicates that beam dynamics is Hamiltonian and it is fully described with
Hamiltonian $H(\zzeta,s)$. The evolution of the distribution function satisfies Vlasov equation which can be considered as a continuity equation in the phase space
\be
\label{Vlasov_general}
\frac{df}{ds}=\partial_sf(\zzeta,s)+\{f,H\}=\partial_sf+(\nabla f)^TJ(\nabla H)=0,
\ee
where $\{f,H\}=(\nabla f)^TJ(\nabla H)$ is the Poison bracket, $J$ is the unit block-diagonal antisymmetric symplectic matrix, $\nabla$ is
the 6D gradient in the phase space, and superscript $T$ stands for transposition.

Typically the electron bunch can be considered well localized and quasi-monoenergetic. Under this assumption, forces can be well approximated to be linear function in respect
to the phase space coordinates. The corresponding Hamiltonian describing beam dynamics is quadratic then, $H(\zzeta,s)=1/2\zzeta^T\mathcal{H}(s)\zzeta$, $\mathcal{H}=\mathcal{H}^T$. The trajectory of
each electron defines the map in the phase space which can be described with the transform matrix $R(s,s_0)$ for linear beamlines. Then formal solution for Vlasov equation (\ref{Vlasov_general}) can
be presented in the following form
\bea
\label{Vlasov_solution}
&&f\left(\zzeta,s\right)=f\left(R^{-1}(s,s_0)\zzeta,s_0\right),\\
&&\frac{dR(s,s_0)}{ds}=J\mathcal{H}R(s,s_0),\;\;\;\;\;\;\;\;R(s_0,s_0)=I.
\eea

Alternatively, the dynamics of modulated beams can be conveniently described in the spectral domain as illustrated in Ref.~\cite{modulation}.
This representation is particularly useful for description of modulated beams since they are well localized in the spectral domain in case of quasi-monochromatic modulations.
The beam can be fully described with its 6D spectral distribution which evolution is Hamiltonian in case of linear optics
\bea
&&f_\kk(\kk,s)=\int f(\zzeta,s)e^{i\kk^T\zzeta}d^6\zzeta,\\
\label{Vlasov_k}
&&\frac{df_\kk}{ds}=\frac{\partial f_\kk}{ds}+\{f_\kk,H_\kk\}=0,\;\;\;\;H_\kk=-\frac{1}{2}\kk^TJ\mathcal{H}J\kk.\;\;\;\;\;\;\;
\eea
The solution of the spectral Vlasov equation was found in Ref.~\cite{modulation}
\be
f_\kk(\kk,s_0)=f_\kk(R^T(s,s_0)\kk,s_0).
\ee
This solution indicates that the spectral distribution function remains constant along characteristics in the spectral domain
\be
\label{k_transform}
\kk(s)=R^{-T}(s,s_0)\kk(s_0),
\ee
which reduces evolution of each spectral component to linear transform of its modulation wavevector.

\subsection{Boltzmann equation in phase space domain}
\label{sec:Boltzmann}

Synchrotron radiation of
highly relativistic electrons is confined within a small angle in respect to the electron instantaneous velocity, $\Delta \theta\sim1/\gamma\ll1$. Therefore, emission of a photon with angular frequency
$\omega$ mainly results in the reduction of the electron energy by $\Delta E=\hbar \omega$. The change of the electron transverse momenta is on the order of $1/\gamma$ smaller then the change of
the longitudinal momentum and can be ignored in the first order. Assuming that the emission
process is instantaneous in time, one can use Vlasov equation (\ref{Vlasov_general}) to describe electron dynamics between photon emission events. Electron energy loss due to photon emission
results in the collision operator which changes phase space density. Large number of electrons within the bunch and ergodicity property allows one to reduce discrete emission events to continuous change of the
ensemble distribution. As a result, evolution of the electron bunch can be described with the following Boltzmann equation
\bea
\label{Boltzmann}
\frac{df}{ds}&=&\partial_sf+\{f,H\}=C^{\rm ISR}[f],\\
\label{C}
C^{\rm ISR}[f]&=&\int\limits_0\limits^\infty f(E+\hbar \omega)\frac{dN(\omega,E+\hbar\omega)}{d\omega}d\omega-\\
\nonumber
&-&\int\limits_0\limits^\infty f(E)\frac{dN(\omega,E)}{d\omega}d\omega,
\eea
where $C^{\rm ISR}[f]$ is the collision operator for ISR energy loss which describes detailed balance between electron states.
$H$ is the beamline Hamiltonian if synchrotron
radiation is neglected, $(dN/d\omega)d\omega ds$ is the probability for a photon emission within bandwidth $d\omega$ while electron travels
distance $ds$. The probability of photon emission is related to the power spectrum $dP/d\omega$ of a single electron synchrotron radiation \cite{Jackson}
\bea
\label{Pomega}
&&\frac{dN}{d\omega}=\frac{1}{\hbar \omega}\frac{dP}{d\omega}=\frac{1}{\hbar\omega}\frac{\sqrt{3}e^2}{2\pi\rho c}\gamma\frac{\omega}{\omega_c}\int\limits_{\omega/\omega_c}\limits^\infty
K_{5/3}(x)dx,\;\;\;\;\\
&&\omega_c=\frac{3}{2}\frac{\gamma^3c}{\rho}.
\eea
Note that collision operator $C^{\rm ISR}[f]$ describing ISR-induced energy losses is linear in respect to the electron distribution function. This property reflects the fact that ISR is a single particle
effect. Superposition of radiation fields from many particles and influence of the resulting Coherent Synchrotron Radiation (CSR) force on particle dynamics is not included in this model.
Presented model also assumes that photons do not interact with other electrons once they are emitted.

Boltzmann equation described by Eq.~(\ref{Boltzmann}) can be used to determine the main parameters of ISR. For example, one can easily find the overall energy loss and the energy diffusion
\bea
\label{P_ISR}
\left(\frac{d\left\langle{E}\right\rangle}{ds}\right)_{\rm ISR}&=&\int EC^{\rm ISR}[f]d^6\zzeta=-P,\\
%\left(\frac{d\bar{E}}{ds}\right)_{\rm ISR}&=&\int EC^{\rm ISR}[f]d^6\zeta=-P,\\
\label{dE_ISR}
\left(\frac{d{\left\langle\Delta E^2\right\rangle}}{ds}\right)_{\rm ISR}&=&\int (E-\left\langle{E}\right\rangle)^2C^{\rm ISR}[f]d^6\zzeta=2D,\;\;\;\;\;\;
%\left(\frac{d\overline{(E-\bar{E})^2}}{ds}\right)_{\rm ISR}&=&\int (E-\bar{E})^2C^{\rm ISR}[f]d^6\zeta=2D,\;\;\;\;\;\;\;
\eea
where $\left\langle{E}\right\rangle=\int Ef d^6\zzeta$ is the mean electron energy and $\left\langle{\Delta E^2}\right\rangle=\int (E-\left\langle{E}\right\rangle)^2fd^6\zzeta$ is the rms energy spread.
%Subscripts $t$ in the  energy loss $P_t$ and diffusion $D_t$ coefficients indicate that these parameters refer to description which uses time as evolutional variable.
These coefficients can be expressed through the synchrotron radiation power spectrum. 
\bea
\label{Pt}
P&=&\int\limits_0\limits^\infty\frac{dP}{d\omega}d\omega=\frac{2}{3}\frac{e^2}{\rho^2}\gamma^4,\\
\label{Dt}
D&=&\frac{1}{2}\int\limits_0\limits^\infty\hbar\omega\frac{dP}{d\omega}d\omega=\frac{55}{48\sqrt{3}}\frac{\hbar e^2c}{\rho^3}\gamma^7.\;\;\;\;\;\;
\eea
Note that the overall change of beam energy and energy spread are described with Eqns.~(\ref{P_ISR}) and (\ref{dE_ISR}) if the beam is not accelerated and the beamline does not contain elements
having nonzero $R_{6i},\; i=1..6$ beam matrix elements, {\it e.g} high order mode RF cavities introducing energy slews.

\subsection{Boltzmann equation and its solution in spectral domain}
\label{sec:Boltzmann_k}

Boltzmann equation described with Eq.~(\ref{Boltzmann}) --- (\ref{C}) is a complicated integro-differential equation. 
On the other hand, this equation take much simpler form in the spectral domain. The appropriate equation can be found by taking Fourier transform of Boltzmann equation
(\ref{Boltzmann}). Its left-hand side is the same as Vlasov equation (\ref{Vlasov_general}) and, therefore, its Fourier transform reduces to spectral Vlasov equation (\ref{Vlasov_k}). The Fourier transform
of the ISR collision operator results in the corresponding collision operator in the spectral domain. One can find it to be equal to
\be
C_\kk^{\rm ISR}=\int f(\zzeta)e^{-i\kk^T\zzeta}\int\limits_0\limits^\infty (e^{ik_E\hbar\omega}-1)\frac{dN}{d\omega}d\omega d^6\zzeta.
\ee

These integrals can be evaluated under assumption that the electron bunch is quasi-monoenergetic. In this case the ISR spectrum can be assumed to be the same for all the particles and the integral over
photon frequency can be evaluated independently from the integral over the phase space which yields to the spectral distribution function. This assumption is equivalent to neglecting with the radiation
cooling effect which is a valid approximation for high brightness linear machines.

The integral over photon frequencies can be evaluated using spectral distribution of the ISR power (\ref{Pomega}) and the spectral Boltzmann equation takes the following form
\bea
\label{Boltzmann_k}
\ds\frac{df_\kk}{ds}&=&\frac{\partial f_\kk}{ds}+\{f_\kk,H_\kk\}=C_\kk^{\rm ISR},\\
\label{C_k}
C_\kk^{\rm ISR}&=&(-ik_EP^{\rm eff}-k_E^2 D^{\rm eff})f_\kk,\\
\label{Peff}
\ds P^{\rm eff}(\varkappa)&=& \frac{54}{55}\frac{5\varkappa\sqrt{1+\varkappa^2}-3\sinh(\frac{5}{3}{\rm asinh}(\varkappa))}{\varkappa^3\sqrt{1+\varkappa^2}}\,P,\;\;\;\;\;\;\\
\label{Deff}
\ds D^{\rm eff}(\varkappa)&=& \frac{9}{8}\frac{-\sqrt{1+\varkappa^2}+\cosh(\frac{5}{3}{\rm asinh}(\varkappa))}{\varkappa^2\sqrt{1+\varkappa^2}}\,D,\;\;\;\;\;\;\\
\label{param}
\varkappa&=&k_E\hbar\omega_c.
\eea
One can note that Boltzmann equation (\ref{Boltzmann_k}) --- (\ref{C_k}) in the spectral domain has much simpler form compared to Boltzmann equation in the phase space domain (\ref{Boltzmann}) --- (\ref{C}).
ISR collision operator in the spectral domain $C_\kk^{\rm ISR}$ algebraically depends on the spectral distribution function which indicates that different spectral components evolve independently
from each other.
The collision operator depends on two parameters, namely the effective energy loss $P^{\rm eff}$ and diffusion $D^{\rm eff}$
coefficients, as described by Eqs.~(\ref{Peff}) --- (\ref{param}). The effective energy loss term results in the change of each spectral component phase which corresponds to the loss of the mean bunch
energy while absolute amplitude of each component is not affected. On the contrary, the effective energy diffusion term results in the reduction of each spectral component amplitude
which manifests as smearing of harmonic current in the phase space domain.

\begin{figure}[ht]
\includegraphics[width=3.5in,keepaspectratio]{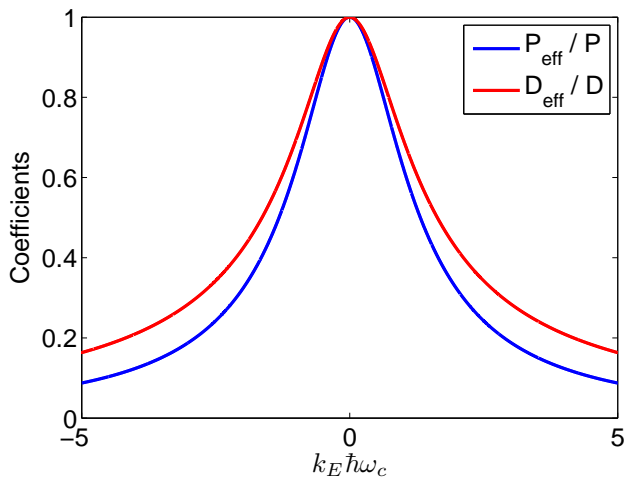}
\caption{(Color online) Dependence of effective energy loss $P^{\rm eff}$ and diffusion $D^{\rm eff}$ coefficients.}
\label{coeffs}
\end{figure}

The effective energy loss and diffusion parameters depend on a single parameter $\varkappa=k_E\hbar\omega_c$. The nonzero energy modulation wavenumber $k_E$ 
indicates presence of energy bands in the phase space distribution
separated by $\Delta E_{bands}=2\pi/k_E$. Electrons passing through the bend emit
photons which reduces electron energy by $\Delta E_{\rm ISR}=\hbar \omega$ in a single emission event.
If all electrons would lose the same energy in a single photon emission event $\left\langle \Delta E_{\rm ISR}\right\rangle=\hbar\omega_c$ then the distribution function would not change
if $\left\langle \Delta E_{\rm ISR}\right\rangle=\Delta E_{bands}$ since all the electrons would move into the same position within the next band and all the bands are equally populated. Then energy 
diffusion would vanish at $k_E\hbar\omega_c=2\pi$ in this hypothetical scenario.
However, photons emitted due to ISR have nearly 100\% frequency distribution, so in reality electrons do not move into the same position within the next energy
band but also acquire some random misplacement which
manifests as energy diffusion. At the same time, the diffusion coefficient at $\varkappa=2\pi$ is much smaller compared to the case of $\varkappa\ll1$ (as illustrated in Fig.~\ref{coeffs})
since a large fraction of electrons do not significantly change their relative position within the corresponding band. 

Boltzmann equation in the spectral domain is a linear hyperbolic equation which can be solved using method of characteristics. The characteristics describing trajectories in the spectral domain are the same
as for Vlasov equation (\ref{k_transform}). This property indicates that transform of modulation wavevector is independent from whether ISR is accounted or not. Taking ISR into account (both overall energy
loss and energy diffusion) results in the change of the harmonic current amplitude and phase along the beamline. At the same time, the optics required to transform initial modulation into required one and
the bandwidth of that modulation are not affected by ISR. This property is highly useful in studies of XFEL seeding schemes since the beamline optics can be designed without accounting for ISR which can be
included at the last stage while estimating reduction of the output bunching factor.

The absolute amplitude of the spectral distribution changes due to effective energy diffusion effect but it is not affected by the overall energy loss. 
One can find solution of linear Eqs.~(\ref{Boltzmann_k}) --- (\ref{C_k}) in the following form which can be used to find smearing of harmonic current due to ISR in any arbitrary linear beamline
\bea
\label{Boltzmann_sol}
\left|\frac{f_\kk\left(\kk(s_f),s_f\right)}{f_\kk(\kk(s_0),s_0)}\right|&=&\exp\left(-\int\limits_{s_0}\limits^{s_f}k_E^2(s)D^{\rm eff}(s)ds\right),\;\;\;\;\;\;\;\;\\
\label{k_tr}
\kk(s)&=&R^{-T}(s,s_0)\kk(s_0).
\eea
There are several properties which follow from Eq.~(\ref{Boltzmann_sol}). First of all, the harmonic current amplitude always reduces along the beamline. This property reflects the diffusive nature of ISR
which smears out small scale variations in the distribution function. Another property states that the rate of modulation smearing is proportional to $k_E^2$. This property reflects the
fact that ISR-induced diffusion is the energy diffusion. Uniform energy distribution is not affected by this effect.
Also note that the integral along the beamline in Eq.~(\ref{Boltzmann_sol}) can be
substituted as a sum of integrals along each element. As a result, the overall attenuation of the harmonic current is equal to the product of attenuations in each dipole.

\subsection{Fokker-Planck approximation}
\label{sec:FP}

Boltzmann equation in the phase space domain (\ref{Boltzmann}) --- (\ref{C})
can be simplified under assumption that the characteristic energy scale of the distribution function $f/\partial_E f$ is much larger than the
characteristic energy of emitted photons
$\hbar \omega_c$. Then the distribution function in the collision operator (\ref{C}) can be expanded and the following Fokker-Planck equation recovered
\be
\label{FP_t}
\frac{df}{ds}=\partial_E(Pf)+\partial_{EE}^2(Df),
%&=&-\partial_{\zeta_6}(P_tf)+\partial_{\zeta_6\zeta_6}^2(D_tf),
\ee
where energy loss $P$ and diffusion $D$ coefficients are described with Eqns.~(\ref{P_ISR}) and (\ref{dE_ISR}). 

Fokker-Planck equation can be solved in the spectral domain in same way as it was done in Sec.~\ref{sec:Boltzmann_k} for Boltzmann equation. However, this analysis is not required since Fokker-Planck
equation is an approximation of a more general Boltzmann equation and its solution can be recovered from a general solution. Boltzmann equation (\ref{Boltzmann}) yields to Fokker-Planck approximation when
the typical energy of emitted photons is much smaller than the characteristic energy scale of the electron distribution function.
This assumption corresponds to the limit of $k_E\hbar\omega_c\ll1$ as discussed in
Sec.~\ref{sec:Boltzmann_k}. Therefore, solution of Fokker-Planck equation (\ref{FP_t}) can be found as a small photon energy limit of the general solution for Boltzmann equation (\ref{Boltzmann_sol}), 
when $D^{\rm eff}=D$.

Smearing of harmonic current depends on the energy modulation wavenumber along the beamline which can be found using transform (\ref{k_tr}). Consider, for example, some beamline for XFEL seeding
scheme. As
discussed in Ref.~\cite{modulation} such a beamline is designed to recover imposed modulation as longitudinal bunching at a given wavenumber, ${\bf k}(s_f)=\hat{\kk}_z k=(0,0,0,0,k,0)^T$. Then the modulation
wavevector at any given position along the beamline can be expressed through the modulation wavevector at the final position and the linear transform matrix as $\kk(s)=R^T(s_f,s)\kk(s_f)$. Then one can find
that the overall attenuation of harmonic current due to ISR-induced energy spread in the Fokker-Planck approximation is equal to
\be
\label{R56}
A\equiv\left|\frac{f_\kk(\kk(s_f),s_f)}{f_\kk(\kk(s_0),s_0)}\right|=\exp\left(-k^2\int\limits_{s_0}\limits^{s_f}DR_{56}^2(s_f,s)ds\right)
%e^{-k^2\int\limits_{s_0}\limits^{s_f}DR_{56}^2(s,s_f)ds}
\ee
This result agrees with qualitative estimate presented in Ref.~\cite{Litvinenko}. The solution shows that the degradation of harmonic current strongly increases in beamlines having large energy dispersion.
Therefore, conventional XFEL seeding schemes utilizing chicanes \cite{HGHG,CHG,EEHG} might be affected by ISR since chicanes are specifically designed to have large energy dispersion $R_{56}$.

%Attenuation factor described with Eq.~(\ref{R56}) was derived using Fokker-Planck approximation of Boltzmann equation which is valid in the regime of $k_E\hbar\omega_c\ll1$. The energy wavenumber of
%modulation can be expressed through the wavelength of final longitudinal bunching and dispersion of the beamline. Then the estimate (\ref{R56}) is valid in the regime
%\be
%\label{FP_limit_E}
%(R_{56}(s,s_f))_{\rm max}\ll\frac{\lambda}{3\pi}\frac{\rho}{\gamma^3\hbar c}.
%\ee

\section{Smearing of bunching in various XFEL seeding schemes}

In this section we calculate attenuation of harmonic current in various beamlines designed for XFELs seeding with microbunched electron beams. In these schemes the beam 
is typically modulated in the wiggler by interacting with external laser at resonant wavelength and modulation at high harmonic is recovered in the following linear beamline
as longitudinal bunching. This dynamics can be conveniently
described in the spectral domain as was illustrated in Ref.~\cite{modulation}. This formalism describes change of the spectral distribution function along characteristics which
can be considered as trajectories of the modulation wavevector in the spectral domain.
The trajectory of each spectral component does not depend whether ISR energy diffusion is neglected or taken into account which follows from Eq.~(\ref{k_tr})
since it is independent on the diffusion coefficient. Taking into account ISR-induced energy diffusion manifests as the next order effect and results in reduction of the harmonic current along
the beamline. Then the exponential factors in Eqs.~(\ref{Boltzmann_sol}) and (\ref{R56}) can be interpreted as attenuation of harmonic current due to ISR-induced energy diffusion.

In our analysis we limit ourselves to modulations which are recovered as longitudinal bunching at required wavelength at
the end of a chosen beamline element. Then the beam modulation at the beginning of the beamline element
is well defined since its transform is described with Eq.~(\ref{k_tr}). From this prospective, the precise mechanism of imposing initial modulation is not important. Therefore,
the attenuation factor for each element will depend on the output bunching wavelength  and beamline parameters.

We will present estimates for the attenuation factor based on the solution
of Fokker-Planck equation described with Eq.~(\ref{R56}) since it is much simpler than solution of Boltzmann equation (\ref{Boltzmann_sol}) and the integrals can be calculated analytically.
This approximation is valid when condition $k_E\hbar\omega_c\ll1$ is held. The energy wavenumber of
modulation can be expressed through the wavelength of final longitudinal bunching and the beamline dispersion.
In this section we will use conventional units used in Beam Physics, particularly we will use relative energy spread instead of
full energy deviation, $\Delta E= \gamma mc^2(\Delta \gamma/\gamma)$. Then condition for validity of Fokker-Planck equation expressed in common units reads as
\be
\label{FP_limit}
(R_{56}(s_f,s))_{\rm max}\ll \frac{2\lambda\rho}{3\lambda_e\gamma^2}=2.39\frac{\lambda [\Angstrom]}{B [T]}\frac{10GeV}{E}\mu m,
\ee 
where $\lambda$ is the output bunching wavelength and $\lambda_e=2\pi\hbar/(mc)\approx 2.43\cdot10^{-12}m$ is  Compton wavelength.

\subsection{Single bend}
First, we estimate attenuation of harmonic current in a single bend which can be the last bend of a more complicated linear beamline. The transform matrix of a bend is described with Eq.~(\ref{Rbend}). Then
attenuation of harmonic current can be found from Eq.~(\ref{R56}) and for a single bend it is equal to
\bea
\nonumber
A_{\rm bend}&=&\exp\left(-k^2\frac{D}{c}\int\limits_0\limits^\alpha R_{56}^2(\theta) d(\rho \theta)\right)=\\
\nonumber
&=&\exp\left(-\frac{55\pi^2}{63\cdot48\sqrt{3}}\frac{\alpha^7\gamma^5}{\alpha_{fine}}\left(\frac{r_e}{\lambda}\right)^2\right)\approx\\
\label{A_bend}
&\approx&\exp\left(-16\left(\frac{E}{\rm 10 GeV}\right)^5\frac{\alpha^7[deg]}{\lambda^2[\Angstrom]}\right),
\eea
where $\alpha$ is the dipole bend angle, $\alpha_{fine}=e^2/(\hbar c)\approx1/137$ is the fine structure constant, and $r_e=e^2/(mc^2)\approx2.818\cdot 10^{-15} m$ is the classical electron radius.

Smearing of harmonic current can be neglected if the exponent argument is smaller than unity. That results in a maximum angle of the bend which can be used in the beamline.
Exceeding this critical value results in a strong smearing of harmonic current due to very strong scaling of attenuation versus bend angle.
The parameters region for importance of ISR debunching agrees with qualitative estimate (\ref{qualitative}). Qualitative estimate
shows the same scaling versus main parameters as in a rigorous analysis but the numerical factor is about a factor of 6 larger since the ISR-induced energy spread was assumed to be acquired in the
beginning of the bend rather than being uniformly distributed along its length.
%
%To estimate validity of Fokker-Planck approximation we note that the largest dispersion $R_{56}(s,s_f)$ occurs when the intermediate position $s$ is the beginning of the bend. Then
%the quantitative estimate (\ref{A_bend}) for attenuation of harmonic current in bend is valid when
%\be
%\alpha^3\ll\frac{4\lambda}{\lambda_e\gamma^2},\;\;\;\;\alpha[deg]\ll0.433\left(\sqrt{\lambda}[\Angstrom]\frac{10GeV}{E}\right)^{\frac{2}{3}}
%\ee

\subsection{Chicane}
Most of the beam based XFEL seeding schemes utilize chicanes \cite{HGHG,CHG,EEHG,EEHG-EEX}. To estimate the effect of ISR in these schemes we consider that the chicane is designed to transform imposed
space-energy modulation as final longitudinal bunching. The detailed analysis of this setup is presented in Appendix~\ref{app:chicane} and it yields to the attenuation factor of
\bea
\nonumber
A_{\rm ch}=&& \exp\left\{-6700\frac{\alpha^7[{\rm deg}] }{\lambda^2[{\buildrel _{\circ} \over {\mathrm{A}}}]}\left(\frac{E}{10 GeV}\right)^5\times \right. \\
\label{A_chicane}
\times && \left.\left[\left(\frac{R_{56}}{\rho\alpha^3}+0.072\right)^2+0.022\right]\right\}.
\eea
One can note that attenuation of harmonic current is stronger for larger chicanes. This result follows directly from the general expression (\ref{R56}) for the attenuation
factor in a general beamline. Typically, the chicane strength scales as $R_{56}\sim\rho\alpha^3$. Then this term becomes dominant compared to other numerical factors and the following
approximate expression for attenuation can be used most of the time.
\be
\label{A_chicane_apr}
A_{\rm ch}\approx\exp\left\{-\frac{\alpha[{\rm deg}] R_{56}^2[\mu m] B^2[T] }{4.7 \lambda^2[{\buildrel _{\circ} \over {\mathrm{A}}}]}\left(\frac{E}{10 GeV}\right)^3\right\}.
\ee
This expression allows one to estimate the effect of ISR in proposed schemes for XFEL seeding with microbunched beam. 

HGHG scheme \cite{HGHG} uses a single modulator and a single chicane. The chicane strength required to achieve maximum bunching is approximately equal to $R_{56}(\Delta \gamma_{mod}/\gamma)
=\mu_{n1}\lambda/2\pi$, where $\mu_{n1}\sim n$ is the first maximum of the n-th order Bessel function,
$J_n^\prime(\mu_{n1})=0$ \cite{modulation}. Then attenuation of harmonic current in HGHG scheme due to ISR can be estimated as
\bea
\nonumber
A_{HGHG}&\approx&\exp  \left\{-5.4\cdot10^{-3}\mu_{n1}^2\left(\frac{E}{10 GeV}\right)^3\times\right.\\
&\times &\left.\left(\frac{\Delta \gamma_{rms}}{\Delta \gamma_{mod}}\right)^2 \frac{\alpha[{\rm deg}] B^2[T]}{(\Delta \gamma_{rms}/\gamma[0.01\%])^2}\right\}.
\eea
Note that degradation of harmonics current does not depend on the wavelength of the output bunching directly. It depends only on the harmonic number used in HGHG scheme since $\mu_{n1}\sim n$ at $n\gg1$.

Similar estimates can be made for harmonic current smearing in EEHG \cite{EEHG} and CHG \cite{CHG} schemes. These schemes utilize two chicanes of different strengths, so the attenuation should be
estimated for both chicanes. The first chicane in both schemes is used to create energy bands in the phase space from the modulation which can be considered as mostly longitudinal bunching,
$k_E(s_f)\gg k_E(s_0)$. Therefore, the estimate (\ref{A_chicane}) can be applied to the first chicane even though it was derived for a chicane which converts spatio-energetic modulation 
into purely longitudinal bunching. Straightforward algebra shows that the ratio of chicane strength and 
modulation wavelength is almost the same for both chicanes in EEHG and CHG schemes,
$(R_{56}/\lambda)_{\rm chicane1}\approx(R_{56}/\lambda)_{\rm chicane2}$. Therefore, both chicanes in those schemes result in the same attenuation
of harmonic current if they are designed using the same dipoles and the difference in their strengths is caused by different drift lengths within the doglegs. Then attenuation of harmonic
current in EEHG scheme can be estimated as
\bea
\nonumber
&A_{EEHG}&\approx\exp  \left\{-10.8\cdot10^{-3}\mu_{n1}^2\left(\frac{E}{10 GeV}\right)^3\times\right.\\
&&\times \left.\left(\frac{\Delta \gamma_{rms}}{\Delta \gamma_{mod2}}\right)^2 \frac{\alpha[{\rm deg}] B^2[T]}{(\Delta \gamma_{rms}/\gamma[0.01\%])^2}\right\},
\eea
where $\Delta\gamma_{mod2}$ is the energy modulation imposed in the second modulator. In this estimate we neglected with energy diffusion due to ISR in the second undulator which modulates the beam. This
effect can be easily accounted under assumption that the undulator energy dispersion is much smaller compared to chicanes. Then the energy modulation wavenumber $k_E=\mu_{n1}/(\Delta\gamma_{mod2}/\gamma)$
is constant along undulator and full attenuation factor for EEHG scheme should be reduced by $\exp(-k_E^2D_{und}L)$, where $L$ is the undulator length and $D_{und}$ is the undulator energy diffusion
coefficient found in Ref.~\cite{ISR:undulator} (the rms energy change by the undulator is $(\Delta\gamma/\gamma)^2=2D_{und}L$).

Similar estimate for CHG seeding scheme yields to
\bea
\nonumber
&&A_{CHG}\approx\exp  \left\{-4.3\cdot10^{7}(M+1)^2\left(\frac{E}{10 GeV}\right)^3\times\right.\\
&&\times \frac{\sigma_z^2[\mu m]}{\lambda^2[\Angstrom]}\left.\left(\frac{\Delta \gamma_{rms}}{\Delta \gamma_{ind}}\right)^2 \frac{\alpha[{\rm deg}] B^2[T]}{(\Delta \gamma_{rms}/\gamma[0.01\%])^2}\right\},
\eea
where $M$ is the compression factor in CHG scheme and $\Delta \gamma_{ind}$ is the energy slew imposed on the beam inside cavity (additional energy at the location of rms bunch length),
and $\sigma_z$ is the bunch length after compression. The estimate for smearing
of harmonic current in CHG scheme indicates that this effect is very strong compared to other seeding schemes. This result comes from the fact that chicanes used in CHG are much stronger than what is
required for HGHG and EEHG schemes. Chicanes in CHG should change relative particle position on the order of the pulse length to provide significant compression 
unlike harmonic generation schemes which require change of electron relative position on the order of the modulation wavelength. Therefore, CHG scheme may be not feasible for seeding XFELs due
to strong ISR-induced smearing of harmonic current.

\subsection{Emittance Exchanger}
Smearing of harmonic current is large in beamlines having large energy dispersion $R_{56}$ as discussed in Sec.~\ref{sec:FP}. ISR-induced attenuation can be reduced by specifically designing beamline without
energy dispersion. This approach implies that conventional schemes for XFEL seeding such as HGHG, EEHG, and CHG cannot be implemented since they rely on the presence of dispersive elements to recover imposed
spatio-energetic modulation as longitudinal bunching. Therefore, one should consider imposing modulation different from modulation in $E-z$ phase plane. For example, one can consider using
Emittance EXchanger (EEX) optics \cite{EEX} to transform transverse modulation into bunching \cite{JMO,EEHG-EEX}.
Properly designed EEX swaps longitudinal and transverse phase spaces of the beam, and therefore,
the energy dispersion of the beamline $R_{56}(s_f,s_0)$ is zero. However, the energy dispersion from the middle of the beamline to the end, $R_{56}(s_f,s)$, is not zero, and attenuation of harmonic current
does not vanish in this setup as follows from Eq.~(\ref{R56}). Additional minimization
of the ISR-induced debunching can be achieved by using doglegs having zero dispersion so that energy modulation is kept small
in the middle of the EEX optics. Zero energy dispersion of the dogleg can be achieved by inserting a quadrupoles triplet in the dogleg drifts so that the entire drift looks like an effective negative drift space
in terms of linear transform matrix \cite{EEX_comp}.

We design the following setup to recover longitudinal bunching using EEX optics. First, the beam is transversally modulated when it passes through the mask. Then the imposed modulation is recovered as
longitudinal bunching in the following EEX optics.
 The dispersion of each dogleg is chosen to be zero to minimize ISR-induced smearing of harmonic current. Such a design requires that the beam is modulated along
$x^\prime$ phase space coordinate in front of the EEX optics. This can be achieved by inserting a focusing lens just in front of the EEX and placing transverse mask in the focus plane of this lens as discussed
in Ref.~\cite{EEX_comp}. The attenuation of harmonic current in this scheme can be calculated using the same algorithm as discussed in Appendix~\ref{app:chicane} and one can find
\be
\label{A_EEX}
A_{\rm EEX}=\exp\left\{-390\frac{\alpha^7[{\rm deg}] }{\lambda^2[\Angstrom]}\left(\frac{E}{10 GeV}\right)^5 \right\} \\.
\ee

\begin{figure}[ht]
\includegraphics[width=3.4in,keepaspectratio]{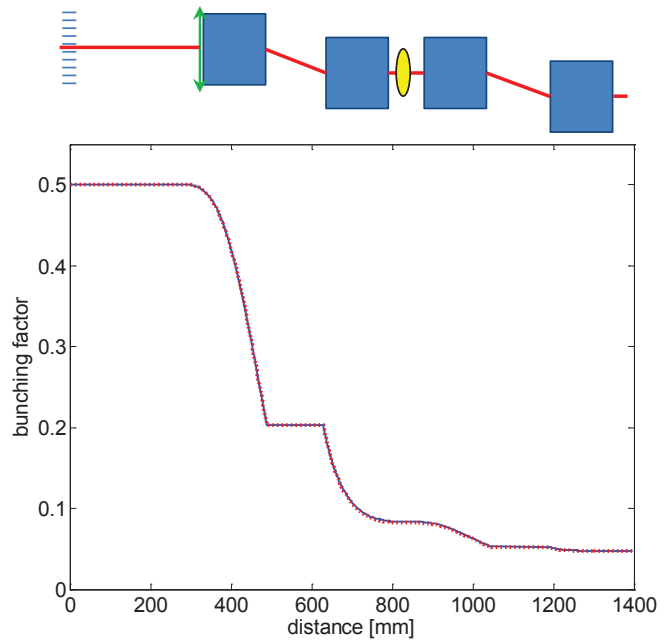}
\caption{(Color online) Attenuation of harmonic current in EEX optics calculated numerically (solid blue line) and analytically (dotted red line).}
\label{fig:EEX}
\end{figure}

We numerically verify the quantitative estimate (\ref{A_EEX}) by simulating EEX optics with a simple particle pushing code. All the beamline elements were considered to be linear and the ISR
was included as a random change of the electron energy at each time step within bends. The results are presented in Fig.~\ref{fig:EEX} for the regime of $k_E\hbar\omega_c\ll1$ in which Fokker-Planck approximation
can be used. The following parameters of the EEX optics were used: beam with mean energy of $E=12GeV$ and rms energy spread of $\Delta\gamma/\gamma=10^{-4}$, normalized transverse emittance $\epsilon_n=0.14$
mm-mrad, beam sizes $\sigma_x=\sigma_y=\sigma_z=100\mu m$ in all dimensions, EEX transforms initial transverse modulation with $23 nm$ wavelength into $0.3\Angstrom$ longitudinal bunching,
EEX dipoles with $B=1T$ magnetic field bend the beam to $\alpha=0.3$ degree angle.
Fig.~\ref{fig:EEX} shows perfect agreement between numerical simulations and quantitative analysis.

Note that attenuation of harmonic current in EEX optics described with Eq.~(\ref{A_EEX}) is similar to attenuation in a chicane having zero energy dispersion $R_{56}=0$ described with Eq.~(\ref{A_chicane}).
This fact can be explained by similarity of these two optics which require the same doglegs having zero dispersion. Moreover, the scaling of attenuation is the same for
both schemes, ${\ln} A\propto -\alpha^7\gamma^5/\lambda^2$, but the corresponding numerical factors are different.
The numerical factor for attenuation of harmonic current in the chicane is on the same order as
for EEX if $R_{56}=0$ and it is an order of magnitude larger compared to EEX if $R_{56}\sim \rho\alpha^3$. Also note that the same scaling holds for a single bend described by Eq.~(\ref{A_bend}).

\section{Discussion}

We developed quantitative approach which allows one to calculate smearing of harmonic current due to ISR-induced energy spread in an arbitrary beamline. The approach is based on a beam representation in the
spectral domain where modulation is represented as a well-localized distribution which evolves along 
the trajectories. The energy diffusion manifests in the spectral domain as a damping operator which allows one
easily calculate attenuation of  harmonic current amplitude along the beamline. The analysis can be further simplified using Fokker-Planck approximation which is valid when the typical energy of
emitted photons is much smaller than the energy bands of modulation.

We applied developed formalism to estimate smearing of harmonic current in various schemes proposed for
XFEL seeding in which imposed modulation is recovered as longitudinal bunching in the following beamline.
We demonstrated that attenuation of harmonic current amplitude increases in beamlines having large energy dispersion $R_{56}$.
At the same time, considered schemes resulted in the same scaling for attenuation of harmonic current, ${\ln} A\propto -\alpha^7\gamma^5/\lambda^2$ as follows from Eqs.~(\ref{A_bend}), (\ref{A_chicane}), and
(\ref{A_EEX}). The only difference
between beamlines came as different numerical factor in front of the scaling. This scaling indicates very rapid increase of ISR-induced debunching effect when either beam energy or the dipole bend angle
increases. As a result, the proposed XFEL seeding schemes should utilize elements with small bend angles. The maximum bend angle can be estimated from the condition that the ISR-induced diffusion does not
reduce the amplitude of harmonic current by more than a factor of 2. Rapid scaling of the attenuation versus bend angle indicates that different beamlines having different numerical factors for attenuation
scaling result in similar critical angles for bends. From this prospective, significant complication of the beamline optics to reduce ISR-induced debunching does not seem practical.

\section{Acknowledgements}
Authors are thankful to D.~V.~Mozyrsky for useful discussions. This work is supported by the U.S. Department of Energy through the LANL/LDRD program.

\appendix
\section{Calculation of harmonic current reduction in chicane}
\label{app:chicane}

In this Appendix we present algorithm for calculating attenuation of harmonic current in chicane which transforms initial beam modulation into longitudinal bunching. However, the same algorithm can be used
to calculate smearing of an arbitrary short-scale modulation in an arbitrary beamline.

Degradation of harmonic current can be described with Fokker-Planck equation (\ref{FP_t}) which describes beam energy diffusion coupled with the 6D phase space transport.
In Secs.~\ref{sec:Boltzmann_k} and \ref{sec:FP}
we assumed phase space variables to be canonical conjugate. However, the variables do not have to be canonical in order to correctly
describe the beam transport since in our analysis of ISR we did not use condition for variables conjugance.
Therefore, one can choose conventional set of variables, {\it i.e.} ${\bf \zzeta}(s)=(x, x^\prime, y, y^\prime, c\Delta t, \Delta\gamma/\gamma)$,
where $x^\prime$ and $y^\prime$ are the particle angles in respect to reference trajectory. 

\begin{figure}[ht]
\includegraphics[width=3.4in,keepaspectratio]{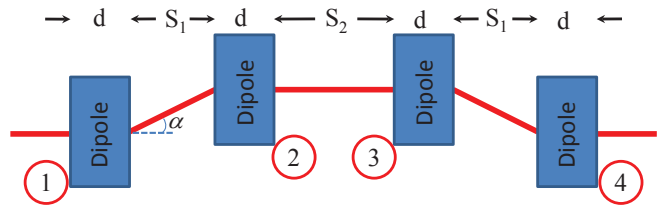}
\caption{(Color online) Schematics of a chicane.}
\label{chicane}
\end{figure}

We describe the chicane as two doglegs separated by a drift space as illustrated in Fig.~\ref{chicane}. To simplify our analysis, we consider hard edge bends. Then particle motion
in $y$-plane is decoupled from motion in other phase space planes and it is represented by a simple drift. Therefore, energy diffusion affects motion only in $x$ and $z$ phase space
planes and the system dynamics can be adequately described in 4D phase space.

The transform matrix of the chicane shown in Fig.~{\ref{chicane}} is equal to
\bea
\nonumber
R_{\rm chicane}&=&R_{\rm flip}R_{\rm dogleg}R_{\rm flip}R_{\rm drift}R_{\rm dogleg}=\\
&=&
\left[
\begin{array}{cccc}
1 &2L+S_2 &0 &0\\
0& 1& 0& 0\\
0& 0& 1& 2\xi\\
0& 0& 0& 1
\end{array}
\right],
\eea
\bea
\nonumber
&&R_{\rm drift}=\left[
\begin{array}{cccc}
1 &S_2 &0 &0\\
0& 1& 0& 0\\
0& 0& 1& 0\\
0& 0& 0& 1
\end{array}
\right],
R_{\rm flip}=\left[
\begin{array}{cccc}
-1 &0 &0 &0\\
0& -1& 0& 0\\
0& 0& 1& 0\\
0& 0& 0& 1
\end{array}
\right],\\
&&R_{\rm dogleg}=\left[
\begin{array}{cccc}
1 &L &0 &\eta\\
0& 1& 0& 0\\
0& \eta& 1& \xi\\
0& 0& 0& 1
\end{array}
\right].
\eea
Here the dogleg parameters can be found in the ultra relativistic limit as
\bea
&&L=\frac{2d\cos\alpha+S_1}{\cos^2\alpha},\\
&&\eta=\frac{S_1+2d\cos\alpha-(2d+S_1)\cos^2\alpha}{\sin\alpha\cos^2\alpha},\\
&&\xi=\frac{2d\sin\alpha\cos\alpha-2d\alpha\cos^2\alpha+S_1\sin^3\alpha}{\sin\alpha\cos^2\alpha},
\eea
where $\alpha$ is the bend angle of the dipole with parallel pole faces, $d$ is the dipole length, $S_1$ is the distance between bends in each dogleg, and $S_2$ is the distance between two doglegs.

Finding attenuation of harmonic current requires knowledge of the modulation wavevector inside each bend. First we find the modulation wavevector at the edge of each dipole using
Eq.~(\ref{k_tr}) which describes its transform along the beamline. We consider the output modulation of the beam to be longitudinal bunching, ${\bf k}(4)=2\pi/\lambda\times[0,0,1,0]^T$.
Then the modulation wavevector at positions 1, 2, and 3 along the beamline can be found as
\bea
&&{\bf k}(1)=R_{\rm chicane}^T{\bf k}(4)=2\pi/\lambda[0,0,1,2\xi]^T,\\
&&{\bf k}(2)=R_{\rm dogleg}^{-T}{\bf k}(1)=2\pi/\lambda[0,-\eta,1,\xi]^T,\\
&&{\bf k}(3)=R_{\rm drift}^{-T}{\bf k}(2)=2\pi/\lambda[0,-\eta,1,\xi]^T,\\
&&{\bf k}(4)=2\pi/\lambda[0,0,1,0]^T.
\eea

The wavenumber of modulation inside each bend can be found using transform matrix of the bend from its edge to some intermediate plane
\be
\label{Rbend}
R_{\theta}=\left[
\begin{array}{cccc}
\cos\theta & \rho\sin\theta &0 &\rho(1-\cos\theta)\\
-\sin\theta/\rho& \cos\theta& 0& \sin\theta\\
-\sin\theta& \rho(\cos\theta-1)& 1& \rho(\sin\theta-\theta)\\
0& 0& 0& 1
\end{array}
\right],
\ee
where the intermediate angle inside the bend is defined as $\theta$ to distinguish it from the overall bend angle $\alpha$ of the dipoles. Then the modulation wavevector in each bend
\bea
&&{\bf k}^{(1)}(\theta)=R_\theta^{-T}{\bf k}(1),\\
&&{\bf k}^{(2)}(\alpha-\theta)=(R_{\rm flip}R_\theta R_{\rm flip})^{T}{\bf k}(2),\\
&&{\bf k}^{(3)}(\theta)=(R_{\rm flip} R_\theta R_{\rm flip})^{-T}{\bf k}(3),\\
&&{\bf k}^{(4)}(\alpha-\theta)=R_\theta^{T}{\bf k}(4).
\eea

Using Eq.~(\ref{Boltzmann_sol}) in the limit of $k_E\hbar\omega_c\ll1$ one can find attenuation of the harmonic current in each dipole. Note that we use of $\Delta\gamma/\gamma$ instead of $\Delta E$ as an 
independent energy variable and the bend
angle $\theta$ instead of $s$ as the path length variable in Fokker-Planck equation. This choice of variables changes the diffusion coefficient to $D_\theta=\rho/(\gamma mc^2)^2 D$
and the energy modulation wavenumber to $k_{\Delta \gamma/\gamma}=\gamma mc^2\, k_E$.
The overall reduction in the modulation amplitude is equal to the product of attenuation factors in each element. Calculating appearing integrals and considering the limit of small bend angles, $\alpha\ll1$,
one obtains the following expression for attenuation of harmonic current in the chicane

\bea
\nonumber
A_{\rm chicane}&=&\exp\left\{{-\sum\limits_{i=1}\limits^4 \int\limits_0\limits^\alpha \left(k^{(i)}_{\Delta \gamma/\gamma}(\theta)\right)^2D_\theta d\theta}\right\}\approx\\
\nonumber
&\approx&\exp\left\{-6700\frac{\alpha^7[{\rm deg}] }{\lambda^2[{\buildrel _{\circ} \over {\mathrm{A}}}]}\left(\frac{E}{10 GeV}\right)^5\times \right. \\
 &\times& \left.\left[\left(\frac{R_{56}}{\rho\alpha^3}+0.072\right)^2+0.022\right]\right\},
\eea
where $R_{56}\equiv2\xi$ is the chicane strength.

%\end{multicols}

\end{document}